\documentclass[aps,amsmath,twocolumn]{revtex4}
\usepackage{graphics, setspace,epsfig,color,subfigure}
\usepackage[letterpaper, dvips,width=7.5in,height=8.5in,includemp=false]{geometry}
\usepackage[vcentermath]{}

\newcommand{\be}{\begin{equation}}
\newcommand{\ee}{\end{equation}}

\begin{document}

\title[title]{A tale of two skyrmions:  the nucleon's strange quark content in different large $N_c$ limits}
\author{Aleksey Cherman and Thomas D. Cohen
  }
\affiliation{Department of Physics \\
University of Maryland\\College Park, MD 20742}


\begin{abstract}
The nucleon's strange quark content comes from closed quark loops,
and hence should vanish at leading order in the traditional large
$N_c$ (TLNC) limit.   Quark loops are not suppressed in the
recently proposed orientifold large $N_c$ (OLNC) limit, and thus
the strange quark content should be non-vanishing at leading
order.  The Skyrme model is supposed to encode the large $N_c$
behavior of baryons, and can be formulated for both of these
large $N_c$ limits. There is an apparent paradox associated with
the large $N_c$ behavior of strange quark matrix elements in the
Skyrme model. The model only distinguishes between the two large
$N_c$ limits via the $N_c$ scaling of the couplings and the
Witten-Wess-Zumino term, so that a vanishing leading order
strange matrix element in the TLNC limit implies that it also
vanishes at leading order in the OLNC limit, contrary to the
expectations based on the suppression/non-suppression of quark
loops.  The resolution of this paradox is that the Skyrme model
does not include the most general type of meson-meson interaction
and, in fact, contains no meson-meson interactions which vanish
for the TLNC limit but not the OLNC.  The inclusion of such terms
in the model yields the expected scaling for strange quark matrix
elements.
\end{abstract}

\maketitle

During the past two decades there has been an extensive
experimental program to study strange quark matrix elements of
the nucleon \cite{exp}.  They are of interest in large measure
because they are sensitive to physics clearly beyond the naive
quark model --- they are nonzero only due to closed strange quark
loops. Thus they are an ideal way to explore an important
theoretical issue: the distinction between two variants of the
large $N_c$ limit of QCD.  In this paper we focus on strange
matrix elements in Skyrme models\cite{Skyrme}, which are chiral
soliton models often justified by appeals to large $N_c$
QCD\cite{WittenOrig,ANW}. Attempting to understand the $N_c$
scaling of strange matrix elements in the context of Skyrme
models raises an apparent paradox which this paper resolves.

The traditional method for generalizing QCD to many
colors\cite{'tHooft, WittenOrig} treats the quark as being in the
fundamental representation of $SU(N)$.  We will refer to this
approach as the 't Hooft (or ``traditional'') large $N_c$ (TLNC).
Recently, an alternative method --- dubbed the ``orientifold
large $N_c$'' (OLNC)  limit\cite{ArmoniShifman,Armoni03,
ArmoniPRL} --- for generalizing to large $N_c$ has been proposed,
where quarks are taken to be in a two-index representation of
color. The principal theoretical motivation for studying this
limit was the connection of one flavor QCD in this limit to large
$N_c$ supersymmetric Yang-Mills theory; this allows one to exploit
powerful mathematical tools in the analysis of one-flavor QCD.
However, there is an important connection to phenomenology: for
$N_c=3$ the anti-symmetric representation is isomorphic to the
fundamental representation.

The fundamental difference between the two approaches is that the
TLNC limit suppresses quark loop effects while the OLNC does not.
Quarks are double-color-indexed objects in the OLNC limit and
scale in essentially the same was as gluons; all planar diagrams
are leading order. Thus, mesons in the OLNC limit scale with $N_c$
in the same way as glueballs\cite{Veneziano} which is distinct
from the scaling in the TLNC limit:
\begin{eqnarray}
\Gamma_n & \sim & N_c^{2-n} \; \; \; \; \; \, {\rm (OLNC)}, \nonumber \\
\Gamma_n & \sim & N_c^{1-n/2} \; \;  \; {\rm (TLNC)}\, ,
\label{meson-scale}
\end{eqnarray}
where $\Gamma_n$ is a generic  n-meson vertex.  In effect, there is
a rule to convert the generic scaling from the TLNC limit to the
OLNC limit, namely, the substitution $N_c^k \rightarrow N_c^{2 k}$.

An obvious consequence of the scaling of Eq.~(\ref{meson-scale})
is on Skyrme models.  The $N_c$ scaling in such models is the result of
the $N_c$ scaling of the parameters in a model. If one alters the
scaling of the parameters of a Skyrmion in the TLNC limit through
the generic replacement $N_c^k \rightarrow N_c^{2k}$ one finds that the
mass of the Skyrmons in the OLNC limit scales as $M \sim N_c^2$.
As shown in refs.~\cite{Bolognesi,ChermanCohen} the $N_c$ scaling of {\it all} generic
properties of the baryon (mass, couplings, cross-sections, etc.) in
the OLNC limit is consistent with the nucleon behaving as a
Skyrmion.  The consistency of this description is made even stronger due to
Bolognesi's observation\cite{Bolognesi} that the coefficient of the
Witten-Wess-Zumino term in the OLNC limit is $N_c(N_c -1)/2 \sim N_c^2$, while in the TLNC limit it is $N_c$\cite{WittenGlobal}.

The consistency of the Skyrme model with large $N_c$ QCD is
deeper than merely showing that all of the generic $N_c$
scaling rules apply; spin and flavor play an essential role.  The
hedgehog structure of the classical solution to the Skyrme model imposes correlations between spatial directions and isospin. These correlations impose relations between certain observables computed at leading order in the
collectively quantized Skyrmions which are independent of the
details of the Skyrme Lagrangian \cite{AdkinsNappi}.  These
relations in all Skyrme-type models encode an emergent symmetry
of QCD --- a contracted SU(2 $N_f$) symmetry where $N_f$ is the number of
flavors.  These rules follow solely from the fact that the
pion-nucleon coupling constant diverges at large $N_c$ while the
pion-nucleon scattering amplitude is finite due to
unitarity\cite{GervaisSakita,DashenManohar,DashenJenkinsManohar}. Since this condition holds for both the
TLNC limit ($g_{\pi N N} \sim N_c^{1/2}$) and OLNC limit ($g_{\pi
N N} \sim N_c$)  the contracted SU(2 $N_f$) spin-flavor symmetry
must emerge in both variants of the large $N_c$ limit of QCD.

To begin, let us focus on {\it the} Skyrme model, {\it i.e.}
Skyrme's original model\cite{Skyrme}, but generalized to three flavors
so that the question of strangeness is relevant. The action for
the model is  \be S=\int \, d^4 x \, \left ( \frac{f_{\pi}^2}{4}
\mathrm{Tr}(L_{\mu} L^{\mu})
    + \frac{\epsilon^2}{4} \mathrm{Tr}([L_{\mu},L_{\nu}]^2) \right
    ) +
    S_{WWZ}
    \label{SK}
\ee
 where the left chiral current $L_\mu$ is given by
$L_{\mu} \equiv U^{\dagger}
\partial_\mu U$, with $U \in SU(3)_f$ \cite{Skyrme,ANW};
$S_{WWZ}$ is the well-known Witten-Wess-Zumino (WWZ) term, the addition of which is necessary for the Skyrme model to respect the symmetries of QCD\cite{WittenGlobal, WittenCurrent}.
The $U$ field can be written as $U = \exp \left ( i \vec{\tau}
\cdot \vec{\pi}/f_\pi \right )$ where $\vec{\pi}$ is the pseudoscalar meson field, and $\vec{\tau}$ is a vector composed of the first three
Gell-Mann matrices, $\vec{\tau}\equiv (\lambda_1, \lambda_2,
\lambda_3)$.  From the scaling rules in Eq.~(\ref{meson-scale}), it is
apparent that $f_\pi \sim \epsilon \sim N_c^{1/2}$ for the TLNC limit,
while for the OLNC limit the scaling is $f_\pi \sim \epsilon \sim
N_c$. The only way that $N_c$ enters is through the parameters
$f_\pi$ and $\epsilon$, and through the Witten-Wess-Zumino
term\cite{ChermanCohen, Bolognesi}.  To show the $N_c$ dependence of the parameters
in an explicit form, we can write
\begin{eqnarray}
f_\pi  &=& \sqrt{N_c} \, \, \overline{f_\pi} \; \; \; \; \; \; \;
\; \; \; \; \; \; \; \; \; \; \; \; \; \epsilon = \sqrt{N_c} \, \,
\overline{\epsilon}  \; \; \; \; \; \; \; \; \; \; \; \; \; \; \; \;{\rm (TLNC)} \nonumber \\
f_\pi &=&\sqrt{ \frac{N_c (N_c-1)}{2} }\, \, \overline{f_\pi} \;
\; \; \; \; \; \epsilon =  \sqrt{ \frac{N_c (N_c-1)}{2}} \, \,
\overline{\epsilon} \; \;   {\rm (OLNC )} \nonumber
\end{eqnarray}
where the barred quantities do not depend on $N_c$.  This implies that the action can be written as
\begin{equation}
S  = N_c \, \, \overline{S} \; \; {\rm (TLNC)} \; \; \; \;
S = \frac{N_c (N_c-1)}{2} \, \, \overline{S} \; \; {\rm
(OLNC )}  \label{S-scale}
\end{equation}
with $\overline{S}$ independent of $N_c$ and of the  same form for
both the TLNC limit and OLNC limit. The choice of the form
$\sqrt{ {N_c (N_c-1)}/{2} }$ rather than $N_c$ for the scaling of
the parameters ensures that the Witten-Wess-Zumino term scales in the
same way as the rest of the system, and is related to the fact that the baryon consists of $N_c(N_c-1)/2$ quarks in the OLNC limit\cite{Bolognesi}.

This leads to an apparent paradox. In general, when a system is in
the semi-classical regime, the size of a prefactor
multiplying the action plays two roles: i) It controls the
convergence of the semi-classical expansion, and ii) specific
powers of the prefactor act as multiplicative factors for
particular observables.  Thus, when $N_c$ is large enough to
justify the neglect of subleading effects in both $1/N_c$
expansions, the only effect of going from the TLNC limit to the
OLNC limit for the Skyrmion is to make the replacement $N_c
\rightarrow N_c (N_c-1)/{2} $ in multiplicative factors
for the various observables.

This is a surprising result, because it appears to leave no room for the effects of the different behaviors of quark loops in the two large $N_c$ limits. At leading order, quark loops are suppressed in the TLNC limit, while not being suppressed in the OLNC limit.  Thus, one would generically expect that strange quark matrix elements should scale as $N_c^2$ in the OLNC limit (that is, with leading order scaling), while in the TLNC limit they should be zero at leading order (that is, they should scale as $N_c^0$, one order below leading).  However, given the simple replacement rule above, it appears that the Skyrme model must predict strange quark matrix of the same scale, $N_c^0$, for both large $N_c$ limits.  The paradox is how to reconcile the expectations for the scaling of strange quark matrix elements from \emph{a priori} quark loop effect considerations with the apparent Skyrme model results.

The resolution would be trivial if the TLNC limit of the Skyrme
model had a leading order contribution to strange quark matrix
elements. While this is counter to our expectations, calculations
of strange quark matrix elements of the nucleon in assorted variants of
Skyrme models have larger typical values than for other
models on the market\cite{exp}.  Since the calculations do not
include {\it any} explicit $1/N_c$ corrections, the very fact that the
results are non-zero seems to suggest that the leading order term
does survive.  However, a careful analysis shows that the strange
quark matrix elements of the nucleon in the Skyrme model {\it
are} zero at leading order in a systematic expansion around the
TLNC limit. While there are no {\it explicit} $1/N_c$
corrections in the existing calculations based on collective quantization,
there are {\it implicit} effects which are subleading in $1/N_c$
and which account for the entire result.

To illustrate this, consider the nucleon's strange scalar matrix
element at zero momentum transfer for a Skyrmion in the exact
SU(3) flavor limit. It is convenient to analyze this matrix element as a fraction, denoted $X_s$, of the total scalar matrix elements
of the three light flavors:
\begin{equation}
X_s \equiv \frac{ \left \langle N \left | \overline{s} s -\langle
\overline{s} s \rangle_{\rm vac}  \right |N \right \rangle} {
\left \langle N \left | \overline{u} u  + \overline{d} d +
\overline{s} s -\langle \overline{u} u+ \overline{d} d +
\overline{s} s \rangle_{\rm vac}  \right |N \right \rangle} \; ,
\label{X}
\end{equation}
where $| N \rangle$ represents the nucleon state and the
quantities with subscript ``vac'' indicating a vacuum subtraction.  For the exact
SU(3) limit, $X_s$ can be computed via collective quantization,
with collective quantum variables specified by an SU(3) rotation $A$
on the standard classical static hedgehog.  That is, $U$ is given by $U=A^\dagger U_h A$
with the hedgehog Skyrmion defined as $U_h \equiv \exp \left (i \hat{r} \cdot \vec{\tau} f(r)
\right)$; the profile function $f(r)$ is determined by minimizing the energy
subject to the condition that the system has unit winding number.  A standard calculation for $X_s$ in the Skyrme model
\cite{Nappi} gives
\be
X_s  =  \frac{1}{3}\langle N \left| 1- D_{8 8}\right | N\rangle = \frac{1}{3}\int dA \; \psi^*_{N}(A)\left(1 - D_{8 8} \right
) \psi_{N}(A)
\label{strangeElement}
\ee
where $dA$ stands for the Haar measure for SU(3), $D_{8 8}=\frac{1}{2} \textrm{Tr}\left[ \lambda_8 A \lambda_8 A^{\dag}\right]$ (which is an SU(3) Wigner D-matrix), and $\psi_{N}(A)$ is the
collective wave function for the nucleon---{\it i.e.,} an
appropriately normalized SU(3) Wigner D-matrix.

As was discussed in another context, tracing the $N_c$ dependence
cleanly requires that the calculation be done with the
coefficient of the WWZ term having an arbitrary explicit $N_c$
dependence\cite{Cohenpenta}.  This implies that the nucleon lies
in a representation that is the generalization\cite{CohenLebed}
of the octet for arbitrary $N_c$.  The generalized representation
``$8$'', which at $N_c=3$ corresponds to the familiar octet, is
specified by $(p,q) = \left ( 1,\frac{N_c-1}{2} \right )$ for the
TLNC limit. The evaluation of Eq.~(\ref{strangeElement}) for
arbitrary $N_c$ can be done straightforwardly with the aid of the
SU(3) Clebsch-Gordan coefficients appropriate for the ``$8$''
representation\cite{Cohenpenta,CohenLebed}.  The result is
\begin{equation}
X_s = \frac{2( N_c +4)}{N_c^2 + 10 N_c +21} = \frac{2}{N_c} + {\cal O}\left(1/N_c^{2}\right) \;  . \label{Xn}
\end{equation}
Thus, $X_s $ goes to zero as $N_c^{-1}$ as $N_c \rightarrow
\infty$. $X_s$ is subleading in a formal $1/N_c$ expansion around
the TLNC limit --- exactly as expected on general grounds.

We note in passing that phenomenological calculations of strange
quark matrix elements are typically done with $N_c=3$ at the
outset in the WWZ term.  This builds in some subleading effects in
$1/N_c$.  For example, calculations of $X_s$ for the exact $SU(3)$
limit \cite{Nappi} gave $7/30$, which is in agreement with Eq.~(\ref{Xn}) for
$N_c=3$.

Thus, it appears that despite our
expectations that strange quark matrix elements of the nucleon in
the OLNC limit are of leading order ({\it i.e.}, ${\cal O} (N_c^2)$), the Skyrme
model of Eq.~(\ref{SK}) has strange quark matrix elements which
are of order $N_c^0$ regardless of whether one is in the
OLNC limit or the TLNC limit. As it happens, this conclusion is
correct, but fortunately it is not the entire story. The fault
lies not in our expectations for the OLNC limit but with the
model: while the direct SU(3) generalization of Skyrme's original
model given in Eq.~(\ref{SK}) does indeed have strange matrix
elements of order $N_c^0$ in the OLNC limit, more general
Skyrme-type models have matrix elements of order $N_c^2$.

We must recall that the general arguments from large $N_c$
QCD do not justify {\it the} Skyrme model in the sense of Skyrme's
original model.   {\it The} Skyrme model does manage to
capture the generic large $N_c$ scaling rules for all observables, as well as the model-independent relations of the contracted SU($2N_f$)
symmetry required from large $N_c$ consistency rules.  Beyond
this, however, values of various couplings predicted by the Skyrme model should be viewed as
model-dependent and thus essentially arbitrary from the point of
view of large $N_c$ QCD.  Indeed, we know {\it a priori} that the
model does {\it not} capture all of the physics at leading order
in $1/N_c$.  For example, {\it the} Skyrme model only has one kind
of meson field, the light pseudoscalar meson field, whereas large $N_c$ in fact has an infinite
number of meson fields.   Even for the light pseudoscalar meson interactions, terms which are
allowed in large $N_c$ QCD are set to zero to make the
calculations tractable; indeed, an infinite number of such terms
are neglected.

Thus, while all terms in {\it the} Skyrme model correctly encode
the leading order large $N_c$ scaling laws for both large $N_c$ limits
(provided the coefficients are scaled properly), the converse is not
true: all terms with the correct leading order scaling behavior
are not included in {\it the} Skyrme model.  The paradox we are
considering is resolved provided there exist terms in Skyrme-type models which,
while absent in {\it the} Skyrme model, are allowed at
leading-order in the OLNC limit and which give rise to strange quark matrix elements.  Such terms
cannot contribute at leading order in the TLNC limit, as they represent quark loop effects.

It is not hard to see how this can happen.  Recall that the principal
difference between the two limits was the suppression of quark
loop effects in the TLNC limit and not in the OLNC limit. One
consequence of this at the level of meson-meson interactions is
that all terms for an underlying SU(3) symmetric theory which
require more than one summation over flavor indices are
suppressed in the TLNC limit by one factor of $1/N_c$ for each
summation beyond the first.  The reason for this is simple: each
distinct sum over flavors for mesons corresponds to distinct quark loops at the quark level.  Each additional quark loop is
suppressed by a factor of $1/N_c$ in the TLNC limit, but not in the
OLNC limit.  Terms with more than one flavor trace that are suppressed in the TLNC limit are known to exist in chiral perturbation theory\cite{GasserLeutwyler}. If terms of
this sort contribute to strange quark matrix elements of the nucleon when included in a Skyrme-type model, the paradox would be
resolved.

To illustrate this idea, consider the effect of the inclusion of one such
term from chiral perturbation theory,
\begin{equation}
{\cal S}'= \int d^4x \, L_4 \, {\rm Tr} \left ( L_{\mu} L^{\mu}
\right ) {\rm Tr} \left(\chi^\dagger U + \chi U^\dagger \right) .
\end{equation}
The coefficient $L_4$ is one of the standard
constants in chiral perturbation theory at order $p^4$, and the
scalar source $\chi$ is taken to be proportional to the quark
masses:
\begin{equation}
\chi = 2 B_0 \,\left ( \begin{array}{ccc} m_u & 0  & 0 \\ 0 &m_d& 0 \\
0 &0& m_s \end{array} \right ),
\end{equation}
where $B_0$ is a constant of proportionality which is order
$N_c^0$. From the discussion above, it is evident that
\begin{equation}
L_4  \sim N_c^0  \; \; {\rm (TLNC)}  \; \; \; \; \; \; \; \;L_4
\sim N_c^2  \; \; {\rm (OLNC)} \label{L4scale}
\end{equation}

Consider a model with an action given by $S_{SK}+S'$. We can probe the strangeness content of this model by considering the strange scalar matrix element at zero momentum transfer, {\it i.e.,} the strange sigma term:
\begin{equation}
\sigma_s \equiv \left \langle N \left | \, m_s \, \left
(\langle \overline{s} s \rangle - \langle \overline{s} s \rangle_{\rm vac} \right
) \right|N \right \rangle = m_s \frac{\partial M_N}{\partial
m_s}\; .
\end{equation}
The second form for $\sigma_s$ is obtained via the
Feynman-Hellmann theorem\cite{CohenGriegel}. Note that this quantity is intimately
related to $X_s$ of Eq.~(\ref{X}) and contains the same
information.

As with the usual Skyrmion, the mass of the nucleon is dominated by the
mass of the classical hedgehog Skyrmion.  The profile
function $f(r)$ is obtained by varying the action subject to the
hedgehog ansatz and imposing a unit winding number.

By standard large $N_c$ rules, the profile function is independent of
$N_c$ at large $N_c$, regardless of whether one studies the TLNC
limit or the OLNC limit.  However, the detailed form of $f(r)$ is
different in the two limits: $S'$ contributes to the leading
order action and hence to the variational equations at leading order in
the OLNC limit, but not in the TLNC limit.

The contribution of the $S'$ term to the mass of the nucleon may be computed
straightforwardly, and from this the Feynman-Hellmann theorem can
be used to compute its contribution to $\sigma_s$, which we
denote $\sigma_s'$:
\begin{eqnarray}
\sigma_s' & = & L_4 \, \,(32 \pi \, m_s \, B_0) \int_0^\infty d r
\, r^2 \left ( f'^2 + \frac{ 2 \sin^2(f)}{r^2} \right ) \nonumber
\\ \sigma_s' & \sim & N_c^0  \; \; {\rm (TLNC)}  \; \; \; \; \; \; \; \;\sigma_s'
\sim N_c^2  \; \; {\rm (OLNC)}
 \label{sigma'}
\end{eqnarray}
where the scaling with $N_c$ follows since everything on the
righthand side of Eq.~(\ref{sigma'}) scales as $N_c^0$ except for
$L_4$; the scaling of $L_4$ with $N_c$ is $N_c^2$, as given in Eq.~(\ref{L4scale}).

The scaling of $\sigma_s'$ in Eq.~(\ref{sigma'}) is precisely as one would have expected from
general arguments involving quark loops in the two limits. Moreover, this behavior is generic. In Skyrme-type models, the inclusion of meson-meson interaction terms the coefficients of which vanish at leading order in the TLNC limit, but not in the OLNC limit, can give rise to strange quark matrix elements of order $N_c^2$ in the OLNC limit.  In the TLNC limit, however, such terms make only subleading (order $N_c^0$) contributions by construction.  This resolves the paradox.

The support of the US Department of Energy through grant DOE-ER-40762-368 is gratefully
acknowledged.

\end{document}